\documentclass[aps,eqsecnum,showpacs,twocolumn]{revtex4}
\usepackage{graphicx}%
\usepackage{amsmath}%
\usepackage{amsfonts}%
\usepackage{amssymb}
\usepackage{bm}
\begin{document}
\draft

\preprint{not yet sub}

\title{Cold atoms in optical lattices: a Hamiltonian ratchet}

\author{ T.S. Monteiro, P.A. Dando, N.A.C. Hutchings, and M. R. Isherwood}

\affiliation{Department of Physics and Astronomy, University College London,
Gower Street, London WC1E 6BT, U.K.}

\date{\today}

\begin{abstract}
We investigate a new type of quantum ratchet which
may be realised by cold atoms in a double-well optical lattice
which is pulsed with unequal  periods. The classical dynamics 
is chaotic and we find the classical diffusion rate $D$ is asymmetric in
 momentum up to a finite time $t_r$.
 The quantum behaviour produces a corresponding asymmetry in the momentum
 distribution which is `frozen-in' by Dynamical Localization provided
 the
break-time $t^* \ge t_r$. We conclude that the  cold
 atom ratchets require  
 $Db/ \hbar \sim 1$ where $b$ is a small deviation
 from period-one pulses.

\end{abstract}

\pacs{32.80.Pj, 05.45.Mt, 05.60.-k}

\maketitle
Cold atoms in optical lattices provide an excellent experimental
demonstration of the phenomenon of {\em Dynamical Localization}
 \cite{Casati}. Dynamical Localization (DL) has been described as the
so-called `quantum suppression of classical chaos'. More precisely, in the usual 
realizations, a periodically-driven or kicked system is associated with classically chaotic 
dynamics for sufficiently strong perturbation. Its average energy grows
 diffusively and without limit, whereas for the corresponding quantum system
the diffusion is suppressed after an $\hbar$-dependent timescale, the 
`break-time' $t^*$. The final quantum momentum distribution is localized
with a characteristic exponential profile. The formal analogy established 
with Anderson Localization  \cite{Fish} forms a key  analysis of this 
phenomenon.

A series of recent experiments on  Cesium atoms \cite{Raizen} gave
a classic demonstration of this effect for atoms 
in driven or  pulsed lattices of sinusoidal form.
The possibility of experiments with asymmetric lattices, in particular with
asymmetric double wells ~\cite{Lucas,Meacher}, leads us to investigate the
 possibility of constructing a `clean' atomic ratchet, where the transport
 results purely from the chaotic Hamiltonian dynamics, with no Brownian or
dissipative ingredients.

 There is already an extensive body of work on Brownian ratchets ~\cite{Brown}, driven
by the need to understand biophysical systems such as molecular motors
and certain mesoscopic systems. Some of this work  encompasses the 
quantum dynamics ~\cite{Reimann}.
 However, to date  there has been very little work on Hamiltonian ratchets. 
One notable exception is the
work by Flach {\em et al} ~\cite{Flach} where the general form of the spatial and temporal 
desymmetrization required to generate transport was investigated.
 To our knowledge,
the  only substantial study of {\em quantum} Hamiltonian ratchets to date,
 however, is the work of Dittrich {\em et al} ~\cite{Ditt} which concluded that
 transport
occurs in mixed phase-spaces. They demonstrated that
overall transport (in their system with periodic boundary conditions) is
zero if starting conditions cover all regions of  phase-space 
(including  stable islands and chaotic regions) uniformly. A key result
was a sum rule which exploited the fact that transport in the chaotic manifold
is balanced by transport in the adjoining regular manifolds (stable islands/tori).
Very recenty ~\cite{Cheon}, it was shown that a kicked map with a
`rocking' linear potential leads to  confinement in the chaotic region
 between a pair of tori which are not symmetrically located about $p=0$.   

Here we propose a new type of Hamiltonian quantum ratchet which, classically,
 is completely
 chaotic. This ratchet is consistent with the rules established
by Dittrich {\em et al} , but relies on a different mechanism:
the transport is due to a non-uniform momentum distribution within 
a completely  chaotic
phase-space. To our knowledge, this is the only example of a clean, non-dissipative
ratchet which does not rely on specific dynamical features such as a particular
set of islands/tori.

 The basic mechanism is as follows: for a repeating cycle
of kicks, of strength $K_{eff}$,
 perturbed from period-one by a small parameter $b$, we find that
the {\em classical} diffusion rates for positive and negative momenta 
($D^+$ and
$D^-$ respectively) are different up to a finite time. Up to this time, an
asymmetry in the classical momentum distribution $N_{cl}(p)$
 accumulates with kick number. 
Beyond this  `ratchet' time, $t_r$, the rates equalize,
we have $D^+ \sim D^- \sim D$ 
(where $D \sim K^2_{eff}/$ is the total diffusion rate),
 no more asymmetry accumulates
and the net classical momentum $<p_{cl}>$ becomes constant.
But  $<p_{cl}^2>$, of course, continues
to grow with time as $\sim D t$. We find that the corresponding quantum
current depends on $t^*/t_r$ : if the break time is too short no
 asymmetry in the quantum $N_{qm}(p)$
accumulates and there is no quantum transport. If $t^*>> t_r$ the
 localization length $L$ becomes large
and the effective quantum momentum asymmetry $\sim <p_{qm}>/L$ decreases. 
We find that $t_r \sim \frac{1}{b^2 D}$.
A quantum ratchet will have the clearest experimental 
signature if $t^* \sim t_r$.
Since $t^* \sim D/ \hbar^2$ , our main conclusion is 
 that optimal cold atom ratchets need
$D b / \hbar \sim 1$ with small $b < 0.2$.

 Here we consider a system with a Hamiltonian given by
 $H=\frac{p^2}{2} + K V(x) \sum_{n,i} \delta(t-nT_i)$.
 We model the double-well ratchet  potential as
$ V(x)=  \{ \sin x + a \sin (2x + \Phi) \}$.
In the usual realization of DL, the Quantum Kicked Rotor
(QKR), the pulses or kicks are equally spaced.
 For the ratchet, we introduce a repeating cycle
of $N$ unequally spaced kicks, ie $T_1,T_2 ..T_N$.

 The time-evolution
operator for the $i-th$  kick of the $n-th$ cycle factorises 
into a free and a 'kick' part $U_{i}=U^{free}_{i}U^{kick}$.
\begin{figure}[ht]
\includegraphics[height=1.5in,width=3.in]{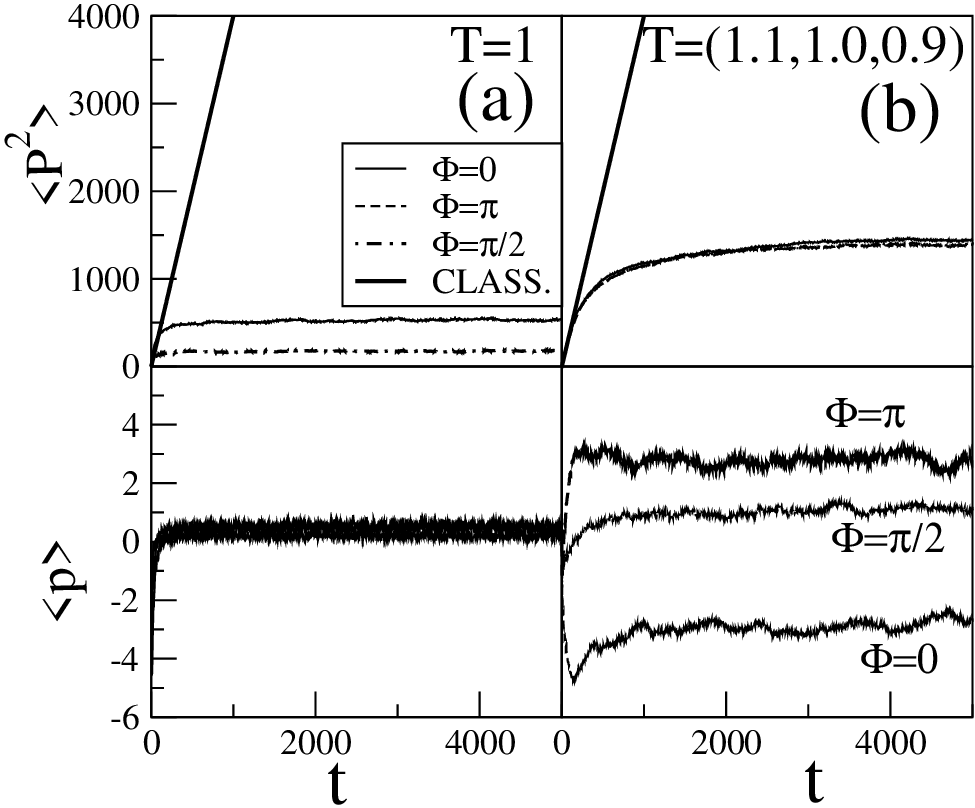}
\vskip 0.5in
\includegraphics[height=1.5in,width=3.in]{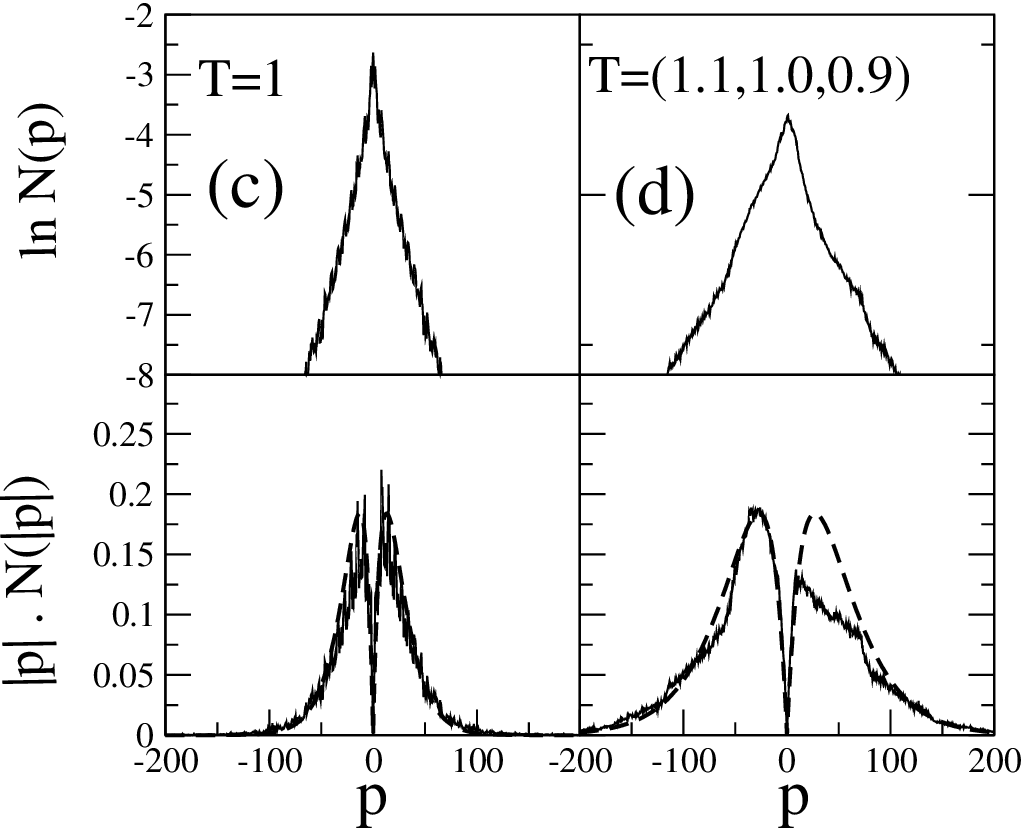}
\caption {Effect of equal and unequal kick spacings on a minimal
uncertainty gaussian wavepacket with $\hbar=1/2$ with initial
 $<p>=0$, for kick strength $K=2$ and $a=1/2$, for different
$\Phi$. (a) Shows evolution of energy and momentum for T=1 (dw-QKR)
(b)Shows evolution of energy and momentum for a repeating $N=3$ cycle of 
kicks with $b=0.1$, hence $T_i=1.1,1.0,0.9$ (cdw-QKR). Figs (a) and (b) show
that the quantum break time is substantially longer for the unequal kicks. 
In (b) we illustrate our finding that the classical
diffusion rate in the chirped case is very well fitted by the lowest order 
approximation $D_0 \sim K^2_{eff}/2 \sim 4$
where $K^2_{eff} = K^2(1+4a^2$) for $D \simeq 2$ or larger.
 They also show that there is no transport
for the equal kick case, but that there is a substantial net momentum
($\sim cst$ if  $t>t^*$) for the cdw-QKR.
Figs. (c) and (d) show corresponding
quantum momentum distributions 
 for the dw-QKR and the cdw-QKR. The figures show that
while the logarithmic distribution takes the usual triangular form
for the dw-QKR, for the cdw-QKR an asymmetric distribution results.
In the lower figure we plot the first moment of the momentum distribution:
(to ease comparison we use $|p|$).
The net momentum is the difference in area between the positive $p$ and
negative $p$ regions. The DL form $|p|. N(|p|)= \frac{|p|}{2L} \exp{-|p|/L}$
(with $L=27.5 \sim 3.5 D_0/ \hbar$ (see fig.3)) is superposed, illustrating that the DL form gives excellent
agreement for all cases for large enough $p$.
 }  
\label{Fig.1}
\end{figure}

 In the 
usual plane wave basis, for a given quasi-momentum $q$,
 the matrix elements of $U_i$ can be shown to take the form:
\begin{equation}
U^{(i)}_{ml}(q) = e^{- i \frac{\hbar T_i (l+q)^2}{2}}
                \sum_s e^{i s \Phi} J_{l-m-2s}(\frac{K}{\hbar})
               J_{s}(\frac{aK}{\hbar})
\end{equation}                  

 where the $J$ are ordinary Bessel functions. The time 
evolution operator for one period 
$U(T=\sum_i T_i)= \prod^{i=N}_{i=1} U_{i}$. In the experiments, 
an important parameter is an effective $\hbar_{eff}=8 \omega_r T$
where $\omega_r$ is the recoil frequency.
In ~\cite{Raizen}, $\hbar_{eff}\sim 1$, so here we have considered
 the range $\hbar=1 \to 1/10$.
 We can use a re-scaled time such that,
without loss of generality, we take
  the average over the periods in each cycle to unity $<T_i>=1$. We define our
`chirped' kicks to mean that we have a repeating cycle of kicks separated by time intervals
 $1+nb, 1+(n-1)b..., 1-(n-1)b, 1-nb$ where $n>0$ is an integer and
$b$ is a small time
increment. $N=2n+1$ for $N$ odd and $N=2n$ for $N$ even. In particular, our
$N=3$ cycle corresponds to a repeating set of kick spacings $T_1= 1+b, T_2= 1 ,T_3= 1-b$
while an $N=2$ cycle corresponds to $T_1=1+b,T_2=1-b$ and so forth.

In Fig.1 we compared the evolution of a quantum wavepacket
 with equal kick times ($T_i=1$)
with a corresponding unequal-kick case with $N=3,b=0.1$. 
Since $V(x)$ in general represents a double well potential, we refer to it
as the dw-QKR to distinguish it from the standard map case
with $V=K \sin x$. For convenience and by analogy with optical terminology,
we refer to the unequal-kick case as the `chirped' or cdw-QKR.
$N=3$ is the minimum number needed to break time-reversal invariance:
 $N=2$ in this system gives no transport. Larger numbers ($N>3$) do give transport
but have the effect of increasing break-times, which, as discussed below,
 we wish to keep small. 

The upper graph in Fig.1 shows that, in both cases,
 the average energy of a classical ensemble of
particles grows linearly with time t ,
ie $<p^2> \sim Dt$.
Neglecting all correlations, it is approximated  by
 $D \sim D_0 \sim K_{eff}^2/2$ where $K_{eff} = K \sqrt{(1+4a^2)}$.
The $\Phi$ dependence appears in neglected correlations which in this case 
appear as products of Bessel functions ~\cite{Hutch}. 
For the standard QKR the corrections appear as oscillations in
$D(K)$, with the maxima corresponding to the $K's$ for which accelerator modes
are important . They were observed experimentally in ~\cite{Raizacc}.
We find that these corrections are  unimportant 
for the cdw-QKR : the lowest order approximation
 gives a good fit from quite low values
 of $K$. The equal-kick case, the
 dw-QKR does need substantial corrections. In fact, since for 
the cdw-QKR phase-space periodicity in momentum is largely destroyed
 and indeed we find no evidence of accelerator modes; the dw-QKR
 is periodic in $p$ and we find 
 strong effects due to accelerator modes.

The {\em quantum} energy follows the classical trend  but
 then saturates to a constant energy  when $<p^2> \sim L^2$,
where $L$ is the localization length, after the break-time $t^*$.
The break-time is substantially longer however
(typically $\sim 3$ times) in the chirped case relative to the dw-QKR. We find that
$t^*$ grows as  $N$. The figure shows that there is 
no net momentum in the $T=1$ case and, asymptotically, $<p> \simeq 0$. 
However, for the chirped case, for $t > t^*$, in general 
we have  $<p> \sim cst$. A meaningful way to quantify the asymmetry 
is a re-scaled momentum $p_L =<p>/L$ which also tends to a constant for $t> t^*$
(eg $p_L \simeq 1/8$ in Fig. 1b for $\Phi=0$.)
Taking $\Phi=\pi$ reverses the symmetry of $V(x)$ and the direction
of motion relative to $\Phi=0$. Intermediate values of $\Phi$ typically
 give $<p>$ within
these extremes.
The  experiments in ~\cite{Raizen} measured the momentum distribution of a cloud of atoms
and observed the  hallmark of DL : a triangular form of the logarithm of the momentum 
distribution $N(p)$. Fig 1(c) (upper) shows this is
 also the case for the dw-QKR. However, for the chirped case,
 the corresponding distribution is evidently asymmetric. 
\vskip .25in
\begin{figure}[ht]
\includegraphics[height=3.5in,width=3.2in]{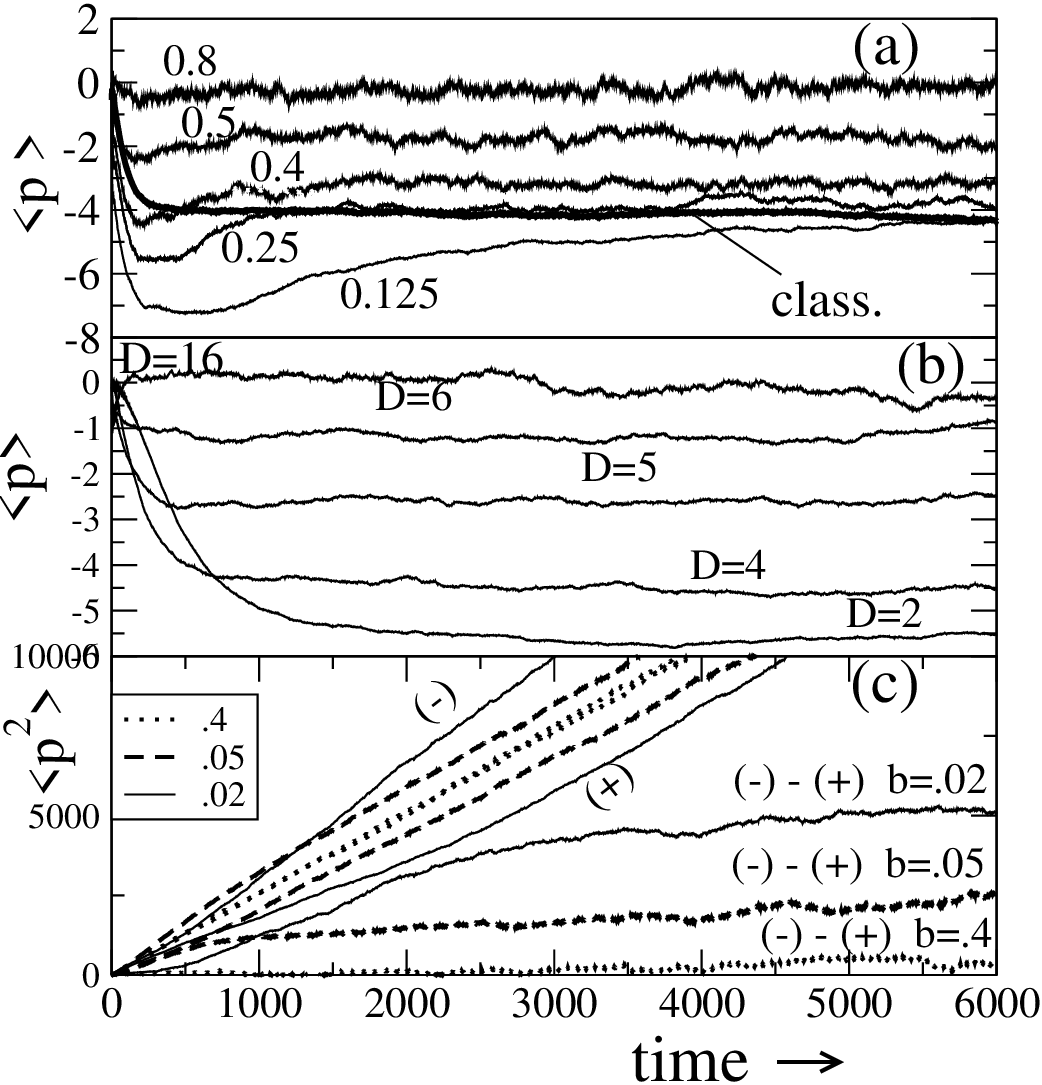}

\caption {Behaviour of the cdw-QKR. The results were obtained from
quantum wavepackets and classical trajectories. Both the classical and
quantum $<p>$ reach a constant value after a finite time ($t^*$ and
$t_r$ respectively).
(a) Graph showing that the quantum ratchet transport is very sensitive to
$\hbar$ if the quantum wavepacket localizes before the classical wavepacket
reaches $t_r$. The values of $\hbar$, ($ 0.8 \to 0.125$) are indicated.
 $K=1.6, a=1/2, N=3$ (hence $D \sim 2.5$) and $b=0.1$. 
The asymptotic  quantum $<p>$ increases with $\hbar$ and `catches' up
with the classical results for $\hbar \sim 0.25$.
(b) Evolution of $<p>$ for a classical wavepacket (500,000 particles with a
gaussian random distribution in $x,p$, of width $\sigma=1.5$ for $b=0.05$ 
but different $D$. 
The classical net momentum increases and then saturates after a classical 
saturation time $t_{r} \sim 1/(D b^2)$.
(c) The effect of different classical diffusion rates ($D^+(t,b), D^-(t,b)$):
$<p^2>$ is evaluated separately for positive and negative
momenta for a cloud of classical particles with $D \sim 2.5$ and different $b$.
 We see that the growth in $<p^2>$
 shows a divergence from linear growth which, for small 
$b \sqrt{<p^2>} $, is  similar in magnitude but opposite in 
sign for the negative and positive components. The $+$ and $-$ indicate
$<p^2>^+ , <p^2>^-$ respectively. One sees that if  $t > t_{r}$ then
  $D^+ \sim D^- \sim D \sim 2.5$.
The lower graphs show $<p^2>^{(-)} - <p^2>^{(+)}$ (for these parameters,
 $D^- > D^+$) and show the $t_{r} \propto 1/b^2$ dependence.  
} 
\label{Fig.2}
\end{figure}
The form for the first moment of 
$N(|p|)$  shows most clearly the source of 
the non-zero average of $p$.
 The dw-QKR is  symmetric. The cdw-QKR distribution, 
shows a substantial asymmetry.
These asymptotic `frozen' distributions are insensitive to changes in
 the starting position in $x$. We have investigated this effect for a range of 
different $K, \hbar$ and $a$.

 In Figs. 2a and 2b we show {\em both} quantum and classical
net momenta $<p>$ increase in magnitude, then saturate to a constant value after 
a finite time $ \sim t^*$ in the quantum case or a  `classical ratchet' timescale $t_r$
in the classical case. But they show other very interesting and surprising
 features.
The quantum dependence on $\hbar$ is shown in Fig. 2a.
The $<p_{qm}>$ are negligible for
$\hbar > 1$ but increase rapidly with decreasing $\hbar$,
up to $\hbar \sim 0.25$. This is important for any
experiment : for these parameters, ($D \sim 2.5, b=0.1$) an experiment with
 $ \hbar_{eff} \sim 0.8$ would show little asymmetry,
 but just halving  $ \hbar_{eff}$ to $ \sim 0.4$ would show
substantial  asymmetry. Beyond $\hbar \sim 0.4$, $<p_{qm}> $ 
is comparable to the saturated classical
value. But since the most experimentally 'detectable'
 ratchet is one which maximizes the asymmetry in Fig.1(d), this means maximizing a re-scaled momentum
$p_L =<p_{qm}>/L$, so there is no advantage in
 reducing $\hbar$ much below $ \sim 0.4$
since $L \sim \hbar^{-1}$.

Fig 2b shows that, for a given $b$ ($b=0.05$ in this graph),
 the classical saturation
time falls with increasing $D$. The fact that $<p_{cl}>$ saturates at all is surprising: after all, the
ensemble of classical trajectories is continually expanding and exploring new phase-space regions
corresponding to higher momenta. Hence specific dynamical
 features such as partial barriers (cantori)
do not account for this ratchet effect, though we find  `scars'
of cantori  do affect the detail
of $N(p)$.

We solve the classical dynamics of the dw-QKR with the usual map ,
$x_{i+1} = x_i +P_i \ ; \ P_{i+1} =P_i - V'(x_{i+1})$ for the $i-th$ kick.
 For the chirped
case, however, we have a 3-kick map (
we consider $N=3$ here but the analysis for $N>3$ is qualitatively
similar). We consider  the first kick of the $n-th$ cycle:
\begin{equation} 
x_{n1}=x_{n0} + P_{n0}(1+b)
\end{equation}
\begin{equation}
P_{n1}= P_{n0} - V'(x_{n0} + P_{n0}(1+b))
\end{equation}
The effect of the chirping is to allow the free evolution to proceed for
an additional distance $\delta_1 =P_{n0} b$. The second kick has 
corresponding $\delta_2=0$ and the third kick has $\delta_3= -P_{n2}b$.
Since $b$ is a small number,
this represents a small perturbation at the start of the evolution of the
cloud and to first order, for the first kick,
\begin{equation}
P_{n1}= P_{n0} - (V'(x_{n0}+P_{n0}) + P_{n0}b V''(x_{n0}+P_{n0}))
\end{equation}
In general, if we work out the change in $<p^2>$ for successive kicks in the standard map
we obtain a diffusion rate $D$ which is the same (after the first few kicks) whether
we average the momentum from $0 \to \infty$ or from $0 \to -\infty$. Both the
zero-th order contribution $K^2/2$ as well as higher corrections (such as those,
for example, involving
correlations between the $i-th$ kick and the $(i+2)th$ kick)
 are even in $p$ and insensitive
to the sign of $p$. 
However, in the cdw-QKR case, if we can make the perturbative expansion above,
we now have correlations which depend on the sign of $p$ and which scale with $b$.

For our classical ratchets, we calculated -separately-  
   $<p^2>^{(-)}$ for those particles with $p<0$ and 
 $<p^2>^{(+)}$ for those with $p > 0$. The results are shown in Fig. 2c 
for $D_0 \sim 2.5$ and different $b$. They
are quite striking : $<p^2>^{(-)}$ and $<p^2>^{(+)}$ separate gradually, more or less
symmetrically, about the line $\sim 2.5 t$, but beyond a certain time they
run parallel to each other and their slopes become equal with
 $D^{(+)} \sim D^{(-)} \sim 2.5$. 
In the lower part of Fig. 2c we plot the actual difference $<p^2>^{(-)}- <p^2>^{(+)}$
for each $b$ since this shows the saturation effect more clearly. 
 
The saturation time $t_r$
  is very 
important since then the classical ratchet speed reaches
 its maximum and for $t > t_r$,  $<p_{cl}> \sim cst$.
We identify it as a point far enough
 from where we can  make  the small angle perturbative
approximation above. This will happen when  $b  \sqrt{<p^2>^{(+)}}  \sim \pi$ for the
positive component and $b \sqrt{<p^2>^{(-)}}  \sim \pi$ for the negative component.
For an order of magnitude estimate of the mean timescale involved,
 we take $b \sqrt{D t}  \sim \pi$.
Hence we obtain
$t_r \sim  \frac{\pi^2}{Db^2}$.

Numerically, we estimate $ t_r \sim  \frac{5}{Db^2}$ which is not inconsistent
with the above. This explains the counter-intuitive behaviour that the larger
 deviation from period-one kicking (ie the larger $b$), give a smaller ratchet effect. 
Though the perturbation
scales as $b$, the time for which it is important scales as $b^{-2}$.
So Fig. 2c shows that $b=0.4$, which saturates at $\sim 10$ kicks, gives $<p> \sim 0$
while $b=0.02$, which saturates at $\sim 5000$ kicks, gives a very large asymmetry.

For the standard map/QKR there is a well-known relation 
between the quantum localization length 
and the classical diffusion constant: $L \sim \frac{\alpha D}{ \hbar}$,
where the constant $\alpha$ was found to be $1/2$
\cite{Shep}. The $N=3, b=0.1$ cdw-QKR takes a modified proportionality
constant, ie $L \sim \frac{3.5D}{ \hbar}$. We have found this by fitting
DL forms to a set of quantum distributions $N(p)$ corresponding to different $D$ and $\hbar$.
For large $p$, the distribution is well approximated by the DL
exponentially localized profile. 

In Fig. 3a we plot a set of calculated $L$ (which range from
 $L \sim 10 - 80$) against $D$ for $ \hbar=1/2$
together with the line corresponding to $L = \frac{3.5D}{ \hbar}$. 
The agreement is excellent.
From  $L^2  \sim Dt^*$ we obtain
$t^* \sim 12 D/\hbar^2$.

In Fig. 3b, for the quantum distributions in Fig 3a,
we have also plotted the net momenta as a functions of $D$, together with their
classical equivalents, obtained from an ensemble of 500,000 classical
 particles. We see that the classical $<p>$ fall monotonically
with $D$, apart from fine structure which we attribute to cantori.
The quantum results however, for low $D$, are much smaller than the 
classical values but increase in magnitude until there is a `cross-over'
 point at $D \sim 3$, after
which they are much closer to the classical values.

\vskip 0.25in
\begin{figure}[]
\includegraphics[height=2.in,width=3.in]{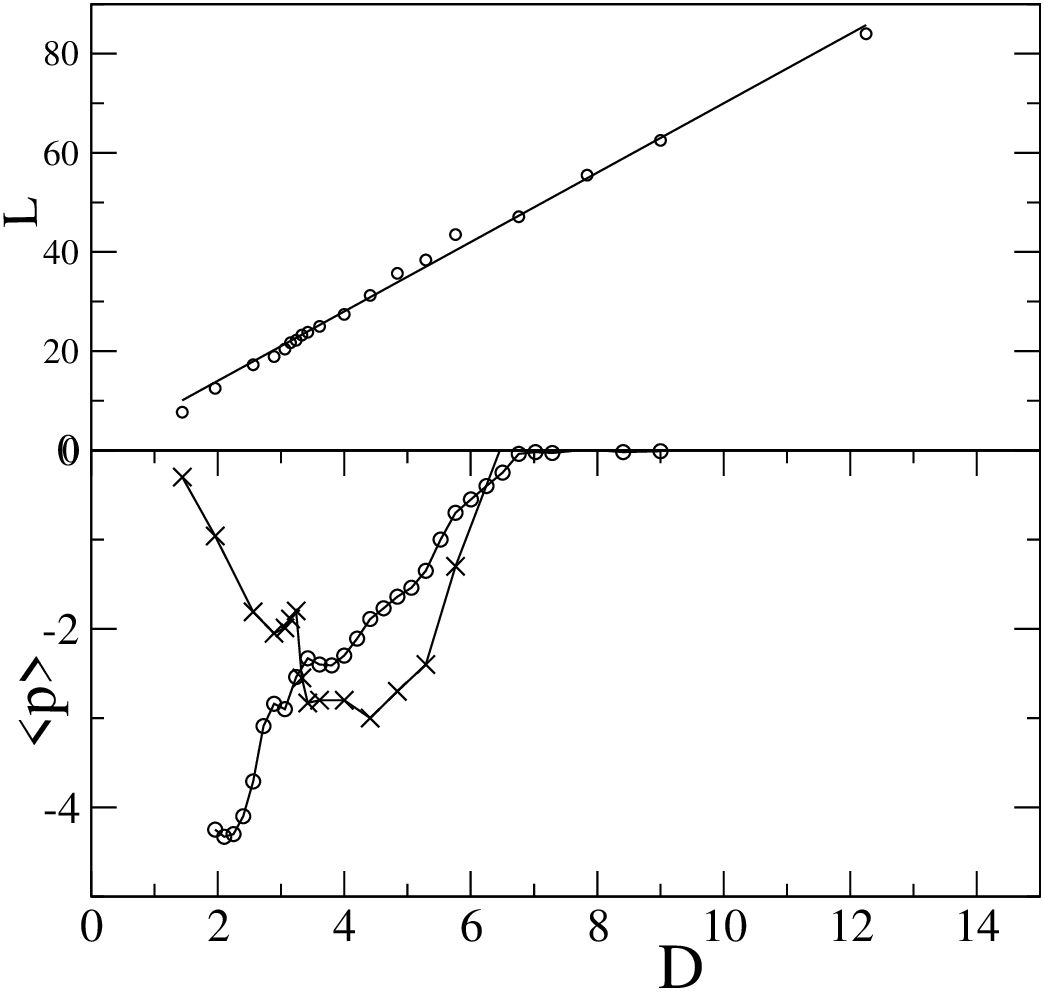}
\caption {(a) Relation between classical diffusion rate $D$ and the
quantum localization length $L$ for the cdw-QKR.
The solid line corresponds to $L = \frac{3.5D}{ \hbar}$.
(b) Average momentum plotted against $D$ for a quantum wavepacket (crosses)
and a classical `gaussian wavepacket' (circles). The graph illustrates the fact that
if the quantum break-time is too short (low $D$) the quantum momentum
is small, but catches up with the classical momentum at $t^* \sim t_r$. }  
\label{Fig.3}
\end{figure}
\vskip 0.25in

  We do not
 expect perfect agreement
with the classical results for $\hbar =1/2$; a cleaner comparison might be obtained
for smaller $\hbar$, but this might be harder to achieve in an experiment.
We  estimate the quantum break-time at the cross-over
$t^* \sim 12D/\hbar^2 \sim 150$ kicks. The ratchet time
 $t_r \sim 5/(Db^2) \sim 160$ kicks.
Such good agreement is somewhat fortuitous, 
since there are larger uncertainties in the timescales.
Nevertheless it does provide us with a useful guide for the best parameters
for an experiment. 

So one of our key results is that the requirement $t^* \sim t_r$ implies that we need
$Db/\hbar \sim 1$. This is a necessary but not sufficient
condition to ensure the best cold-atom ratchet. Clearly the $L$ values should be
experimentally plausible $L \sim 10- 100$, so this places a constraint on $D/\hbar$.
Equally, for a good classical ratchet, $b$ should be small ie $b < 0.2$.

Although the values of $D$ considered may be small
compared with those of the standard map, we find the chirped dw-QKR is 
much more chaotic at a given value of $D$. There are no remaining stable islands
visible on an SOS at the parameters corresponding to Figs 1-3.

In conclusion, we have found a mechanism for a generic, completely chaotic
Hamiltonian ratchet. But much remains
to be done, such as considering the effect of the partial loss of coherence in
the experiment or the effect of the finite duration of the experimental kicks.
Though transport appears as a general sign-dependent correction to the classical
diffusion rate , the role of specific dynamical effects such as desymmetrization of Levy
flights ~\cite{Flach} must be investigated.

We thank Prof. S Fishman for helpful advice. N.H and M.I acknowledge
EPSRC studentships.The work was supported by EPSRC grant GR/N19519.

\end{document}